\documentclass[epj]{webofc}
\usepackage[varg]{txfonts}   
\usepackage{graphicx,color} 
\usepackage{bm}       
\usepackage{amsmath}  
\usepackage{amssymb}
\woctitle{21st International Conference on Few-Body Problems in Physics}
\begin{document}
\title{Exploring the few- to many-body crossover using cold atoms in one dimension}

\author{Nikolaj Thomas Zinner\inst{1}\fnsep\thanks{\email{zinner@phys.au.dk}}}

\institute{Department of Physics and Astronomy, Aarhus University, Ny Munkegade 120, DK-8000 Aarhus C, Denmark
          }

\abstract{
  Cold atomic gases have provided us with a great number of opportunities for studying various physical systems
	under controlled conditions that are seldom offered in other fields. We are thus at the point where one can truly
	do quantum simulation of models that are relevant for instance in condensed-matter or high-energy physics, i.e. 
	we are on the verge of a 'cool' quantum simulator as envisioned by Feynman. One of the avenues under exploration 
	is the physics of one-dimensional systems. Until recently this was mostly in the many-body limit but now 
	experiments can be performed with controllable particle numbers all the way down to the few-body regime. 
	After a brief introduction 
	to some of the relevant experiments, I will review recent theoretical work on one-dimensional quantum systems 
	containing bosons, fermions, or mixtures of the two, with a particular emphasis on the case where the particles are 
	held by an external trap.
}
\maketitle
\section{Introduction}
\label{intro}
Atomic gases cooled to extremely low temperatures are a very rich and 
expanding field of physics where one may explore many different features
of quantum few- and many-body systems in a highly controlled manner \cite{lewenstein2007,bloch2008,esslinger2010,zinner2013}.
In these systems one has accurate control over both geometry and interactions, and 
a hotly pursued feature has been the realization of systems in 
one dimension (1D) \cite{moritz2003,stoferle2004,kinoshita2004,paredes2004,kinoshita2006,haller2009,haller2010,pagano2014}. 
The interactions are also controllable in 1D \cite{olshanii1998} and one may even 
get to the hard-core regime and study the Tonks-Girardeau limit \cite{tonks1936,girardeau1960}
in experiments \cite{kinoshita2004,paredes2004}. In the latest developments it has
now become possible to use 1D setups to reach the limit of small particle 
number \cite{serwane2011} and few-body fermion systems have been realized in 
several experiments in the Jochim group in Heidelberg \cite{zurn2012,zurn2013,wenz2013,murmann2015a,murmann2015b}.
A lot of theoretical research has been devoted to explore few-body physics in
1D and also to try to extrapolate into the many-body regime.
Here I will review
some of these developments with an excusable bias toward some of the work 
performed in our collaboration. Unless otherwise stated I will focus exclusively 
on 1D systems with short-range interactions that are implemented by a Dirac 
delta function. The basic Hamiltonian is 
\begin{equation}
H=\sum_i\left[\frac{p_{i}^{2}}{2m}+V_e(x_i)\right]+\sum_{i<j}V(x_i-x_j),
\end{equation}
where the sums run over the number of particles, $i=1,\ldots,N$, 
$x_i$ is the coordinate and $p_i$ the momentum of the $i$th particle. 
The
external trapping potential $V_e(x)$ is assumed to be the same for all particles.
The mass $m$ is assumed to be the same for all particles in most of the discussion 
below, except for the section on mixtures where I will explicitly 
state when different masses are considered.
The two-body interactions are given by terms of the form $V(x_i-x_j)=g\delta(x_i-x_j)$,
where $g$ is the coupling strength. When I talk about strong interactions 
it is the limit where $1/g\to 0$. For fermions the interactions are typically 
only between pairs of opposite spin projection. However, due to the Pauli principle
enforcing the antisymmetry of pairs with identical spin projection, the interaction
terms in the Hamiltonian above will have no consequence for identical fermion pairs.
For bosons with more than one internal component, one has 
to consider the possibility of interactions between 
the pairs with the same internal components (intraspecies) and pairs with different internal components (interspecies). 
Intra- and interspecies interactions do not have to be the same in general. 
They will all be assumed to 
be given by Dirac delta functions. In the discussions below I will make it clear 
whenever this is important.

\section{Bosonic systems}
A particular focal point of attention in the work on few-boson systems has been the 
transition from weak to strong interactions. In the presence of strong
short-range interactions, a pair of bosons will have a wave function that has to 
vanish when the two particles overlap. As noticed in the famous paper of Girardeau \cite{girardeau1960}, 
this means that the ground state may now be described by a totally antisymmetric fermionic wave function if one 
makes appropriate changes of signs to ensure bosonic symmetry. For instance, 
if you take $N$ identical bosons with no internal degrees of freedom and put them into some 
trapping potential, then the ground state for strong interactions is obtained by
taking the $N$ lowest single-particle eigenfunctions of the single-particle 
Schr{\"o}dinger equation and construction their totally antisymmetry (Slater 
determinant) $N$-body wave function, and finally taking the norm of that to make
sure that the wave function does not change sign under exchange of any two 
particles. 

A number of studies have explored the physics of few-boson systems as the 
interactions are changed from weak to strong using different methods 
such as the Bethe ansatz in a hard-wall trap \cite{batchelor2005,hao2006,oelkers2006}, 
the multi-configurational method \cite{zollner2005,tempfli2008,zollner2008} for 
double-wells and
exact diagonalization \cite{deuret2007,garcia2013a,campbell2014,garcia2014b} or diffusion Monte Carlo \cite{garcia2013b,garcia2014a}
in a harmonic trap. In the 
latter case of a harmonic trap, Girardeau has given a number of 
exact solutions for strong interactions \cite{girardeau2011} that also 
apply for bosons with multiple internal states (spinor bosons). The fact
that the exact solution of the two-body problem in a harmonic trap 
is available \cite{busch1998} has been used to propose analytical $N$-body states 
\cite{brouzos2012}. This is very similar to using Jastrow factors
to incorporate correlations \cite{brouzos2014,wilson2014}.

In recent years, we have explored the physics of two-component Bose systems
in harmonic traps, i.e. similar mass bosons with two internal degree of freedom that
we denote $A$ and $B$. In the limit where the $AA$ and $BB$ interactions are
negligible compared to $AB$ interactions, we found that one may solve 
the three-body problem exactly for strong interactions \cite{zinner2014}. 
The ground state turns out to have an energy that is half an odd
integer times the harmonic trap quantum, and its wave function (which 
is doubly degenerate) displays the precursor of ferromagnetism, i.e. 
the spatial ordering of the particles is either $AAB$ or $BAA$, but {\it never}
$ABA$. On the other hand, one can find excited state in the system that 
{\it only} display the $ABA$ spatial ordering. In our latest 
work \cite{dehk2015a}, we confirm that this kind of fractional 
energy behavior persists for larger systems and we present
analytical results for the four-body state (two $A$ and two $B$ atoms)
for strong interactions. Indeed, already for balanced ten particle
systems with strong $AB$ interactions one can see the perfect ferromagnetic
ordering emerge in the harmonic trap with a profile that approximates 
that expected in the many-body limit \cite{dehk2015a}. Another very
recent study focusing on the strongly interacting limit also finds
ferromagnetic behavior in this parameter regime \cite{pietro2015}.

\section{Fermionic systems}
The solution of the two-component 
Fermi system in 1D in known in the homogeneuos case from the 
nested Bethe ansatz of Gaudin and Yang \cite{gaudin-yang}. However, 
in the case of a non-trivial confining potential no exact solution 
to the $N$-body problem is known for general interaction strength.
Here I will focus on harmonically trapped fermions but stress that 
many of the results and methods can be applied to other external 
confining geometries as well. 
An approach under exploration is to use Bethe ansatz in conjunction 
with a variational principle \cite{rubeni2012} while also the 
multi-configurational Hartree method has been applied \cite{brouzos2013} 
However, at this point it is safe to say that many studies rely in
one way or another on exact 
diagonalization \cite{bugnion2013,gharashi2013,sowinski2013,damico2014,gharashi2014,lundmark2015,gharashi2015,sowinski2014}.
In the strongly interacting limit, some exact solutions are 
in fact known \cite{guan2009,yang2009,cui2014}, although we caution 
that using the spin formalism of for instance Ref.~\cite{guan2009} 
can lead to errors already at the level of four particles \cite{gharashi2013}.
In our recent work on few-fermion physics, we have considered three-body 
problems both for two- \cite{loft2015} three-component systems \cite{artem2014}.
A new variational approach was introduced in Ref.~\cite{loft2015} which 
aims to find an accurate energy and wave function for the three-body 
problem using knowledge of the extreme limits of vanishing and 
infinite interaction strength. This provides a very accurate 
approximation that even extends to mass-imbalanced systems. Most 
recently, we have have considered dynamics of strongly interacting
few-body systems in the case where one has 
time-dependent external traps. This may lead to dynamical engineering
of magnetic quantum states \cite{artem2015f}. A related question is 
how the dynamics depends on Bose of Fermi statistics, where even 
for static traps and weak interactions one sees differences in 
transport properties \cite{chien2012}.

The interesting question of how to get from the few- to the many-body
limit has received much attention lately, in particular for strongly 
interacting particles. Again, Girardeau has written down some exact
solutions to the two-component Fermi system in a trap with 
strong interactions \cite{girardeau2010}
that are simple and beautiful, and have captured much attention. 
While they are of great theoretical use, they have little experimental 
relevance as they are superpositions of states that are adiabatically
connected to experimentally relevant states for large but finite 
interaction strength. This was predicted in Ref.~\cite{volosniev2013}
and has recently been confirmed experimentally in Ref.~\cite{murmann2015b}.
Studies of the few- to many-body limit have been done using diffusion 
Monte Carlo \cite{astrak2013}, exact diagonalization \cite{lindgren2014,deuret2014,levinsen2014}, 
and most recently also 
coupled-cluster methods from quantum chemistry \cite{grining2015}.
A particularly interesting feature is the fact that one may map 
the strongly interacting problem onto a spin model of the Heisenberg type
\cite{deuret2014,volosniev2014,levinsen2014}, but with nearest-neighbor
interactions that now depend on the external confinement and are not constant
in general.
It is interesting to note that the ansatz solution of the strongly 
interacting problem presented in Ref.~\cite{levinsen2014} allows one
to obtain a very accurate expression for the Heisenberg Hamiltonian 
in the case of harmonic confinement. 

In our group we have pioneered
the use of the Lee-Suzuki effective interaction method combined with 
exact diagonalization for 1D systems 
with short-range interactions \cite{dehk2015a,lindgren2014}. It 
provides a large reduction in computational time over traditional 
exact diagonalization with the bare Dirac delta function interaction.
This has allowed us to consider for instance the polaron problem, 
i.e. a Fermi gas of one component (spin up) and a single (impurity) 
particle of the 
other component (spin down), for up to ten particles. Here one see
a very characteristic phase separation with the impurity going to 
the middle of the trap causing a deep indentation in the majority
particle density. This is also seen in the recent study of Ref.~\cite{levinsen2014}
where the many-body limit could be explored analytically using a 
very accurate ansatz wave function.

\section{Mixed systems}
Last but certainly not the least interesting are the mixed systems. 
This should be understood as those systems which are mixtures of 
fermions and bosons that may be either of equal or unequal mass. 
As in all the previous examples, Girardeau has once again 
been one of the first to consider such systems \cite{girardeau2004,girardeau2007}.
In Ref.~\cite{girardeau2004} a very interesting Bose-Fermi mapping was introduced.
However, as we have pointed out in Ref.~\cite{volosniev2013}, these 
mappings are not useful for systems under non-trivial external 
confinement, and have to be applied with great care. While 
for three-body problems with equal mass Bose-Bose (spinor) mixtures one
can indeed obtain the correct results for strong interactions 
\cite{deuret2008}, this does not extend to larger systems using 
Bose-Fermi mappings \cite{volosniev2013}. In Ref.~\cite{fang2011}
a four-body Bose-Fermi mixture with equal mass is considered in 
the strongly interacting regime, and the results are compared 
to a numerical study of the trapped system using the density-matrix
renormalization group (DMRG) method. The Bose-Fermi mapping predicts a 
totally fermionized state in the strongly interacting limit that is 
identical to that obtained in a four-body system of identical (spinless
or spin-polarized) fermions. This latter state has a characteristic four-hump 
structure in the harmonic trap. The DMRG calculations in Ref.~\cite{fang2011}
seem to confirm that this is indeed the strongly interacting ground state. 
However, applying the formalism of Ref.~\cite{volosniev2013} one may confirm 
that this is not the case, and that the density akin to four spinless 
fermions is only found in an excited state. This leads one to suspect
that the DMRG (built on a variational principle) 
may have gotten stuck in an excited state. This is plausible
due to the (quasi)-degenerate manifold of state found in the strongly 
interacting limit \cite{volosniev2013}. This highlights the need for
(semi)-analytical insight into the strongly interacting regime in spite
of the success of the DMRG method over the past decades. 

The problem of mixtures of Bose and Fermi atoms has already an intricate
structure at the level of three atoms \cite{artem2014,harshman2012,barf2015}.
In particular, the strongly interacting limit and the (quasi)-degenerate
manifold of state for large but finite interaction depend delicately on the 
composition, i.e. whether one considers two bosons and a fermion (all of equal mass) 
or vice versa \cite{harshman2012,barf2015}, or one can consider taking three 
altogether different components \cite{artem2014}. In light of this it is a 
very interesting question how can we determine which final state for strong 
interaction will be reach 
given an initial state for vanishing interaction strength? More precisely,
if we adiabatically increase the interacting strength from zero to a 
very large value (effectively infinite), which state is reached? This 
question was explored in great detail in Ref.~\cite{harshman2014}, 
and further elaborated by the same author in Ref.~\cite{harshman2015}. 
Using the symmetries of the problem and general mathematical arguments, 
Ref.~\cite{harshman2015} argues that for the case of four particles 
or less one may solve this issue by using the symmetries of the system. 
On the other hand, when the particle number is larger than four, one can
no longer expect this and must resort to other means. This is an interesting
question that certainly deserves further exploration. 

Another important mixed system is the unequal mass case \cite{mehta2014,artem2015b,dehk2015b,pecak2015,mehta2015}.
Some recent results have demonstrated a phase separation in the light 
component for Fermi-Fermi mixtures \cite{pecak2015}, which can also be seen 
for a single heavy impurity in a light Bose system \cite{dehk2015b}.
In fact, the case of a single impurity in a Bose gas is often called the 
'Bose polaron' problem and is much studied at present in the few-body 
limit \cite{artem2015b,dehk2015b,mehta2015}. In our recent work we 
have explored the dynamics when the boson-impurity interaction is 
suddenly switched on. While the formalism used there works for general 
dimensions, we focused on the 1D case in Ref.~\cite{artem2015b} where
this kind of quench protocol leads to oscillatory behavior in the weakly
interacting regime. In Ref.~\cite{dehk2015b} we present our results on 
the static configuration of the Bose polaron using a new formalism that 
is a hybrid of the hyperspherical adiabatic approach. What is particularly
interesting about this latter work is the fact that one can extrapolate
to large particle numbers with miniscule computational effort. In fact, 
as we argue in Ref.~\cite{dehk2015b} we expect the approximations we 
make to work better for larger systems. By comparison to numerically exact
results \cite{dehk2015a} for systems with ten particles or less we confirm 
that the (semi)-analytical method put forth in Ref.~\cite{dehk2015b} is 
accurate within a few percent already for eight bosons and an impurity.
In addition, this new formalism can handle any trapping potential that 
can be different for the bosons and the impurity, it can tackle any 
mass ratios, and the traps may even be displaced from one another.

\section{Outlook}
As I have argued above, the experimental situation is very promising 
for studying few-body physics in 1D, as well as for gaining knowledge
about how one extrapolates to the many-body regime. I like to call this 
few-body physics in a many-body world. A considerable body of theoretical 
work has been put forth in this area and more is certainly to come. Many
new avenues are being explored and I believe that new aspects of 1D physics 
will be found. I think that an important outstanding question concerns 
the notion of strong in real life. What does 'strong interaction' really 
mean? Above I have used the mathematical definition which means that you take
a coupling constant to infinity. Some recent experimental results \cite{murmann2015b} 
strongly suggest that we understand that limit \cite{volosniev2013}. 
This was done with relatively few particles and I would really like 
to see how things change as we increase the number of particles, preferably
one at a time. I am also very interested in exploring dynamics in 
1D systems with strong interactions such as quantum state transfer \cite{volosniev2014}
and dynamical quantum magnetic effects \cite{artem2015f}. An 
intriguing question that comes to mind in this respect is that of 
spin-charge separation. It is know from 1D Luttinger liquid formalism
that the low-energy excitations of a two-component Fermi gas in 
1D split into charge density and spin density waves, and furthermore that 
these waves have different velocities, hence we get spin-charge 
separation. As far as I know, at this point it has not been unambiguously 
observed in electronic systems such as nanowires. What I would really like
to see is how this effect emerges from a few-body perspective.

\begin{acknowledgement}
I thank my collaborators Artem Volosniev, Amin Dehkharghani, Oleksandr Marchukov, Dmitri Fedorov,
Aksel Jensen, Manuel Valiente, David Petrosyan, Jonathan Lindgren, Christian Forss{\'e}n, 
Jimmy Rotureau, and Hans-Werner Hammer. The experimental group of Selim Jochim in Heidelberg 
is gratefully acknowledged for discussions and for making their data available to us.
This work is supported in part by the Danish Council for Independent Research DFF and the
DFF Sapere Aude program.
\end{acknowledgement}

%
%
%

\end{document}